\documentclass{article}
\usepackage[numbers]{natbib}

\begin{document}
\title{Five Differences Between Ecological and Economic Networks}
\author{Reginald D. Smith \\PO Box 10051\\Rochester, NY 14610\\rsmith@bouchet-franklin.org}
\date{August 27, 2011}

\maketitle

\begin{abstract}
 Ecological and economic networks have many similarities and are often compared. However, the comparison is often more apt as metaphor than a direct equivalence. In this paper, five key differences are explained which should inform any analysis which compares the two. Keywords: complex networks, economic networks, ecological networks, trophic cascades, bullwhip effect
\end{abstract}

\section {Introduction}

The comparison of biological and economic networks has a long history \cite{hist1, hist2, hist3, hist4, hist5, hist6}. Charles Darwin himself admitted that Thomas Malthus was a key influence on his later work on the Theory of Evolution by Natural Selection. In economics, Alfred Marshall famously made some biological analogies to economics in his \emph{Principles of Economics} \cite{marshall1, marshall2, marshall3} such as the similar advantages of diversification amongst organisms in body structure and the division of labor (in particular see book IV, chatper 8). In particular, economic and ecological networks are usually compared by three general features: the structures and functions of their respective networks of interaction, dynamics of competition, and the basic processes of innovation and evolution. Though the comparisons are often generally accurate, many times they drift into metaphor without firm basis and actually ignore profound differences between the two systems in their structures and dynamics. This brief article outlines five differences that stand out and separate the two phenomena.

\section{Five Differences}
\begin{enumerate}

\item	Unit of growth: The basic units of population growth in biology are relatively clear and discrete units from the cell to the elephant. Though mutualist interactions such as those defining lichens can create tightly interacting communities, the unit of analysis is usually firmly on the individual. These discrete units can reproduce asexually or sexually and are typically modeled using population density in a fixed area growing as Malthusian exponential growth (without constraints) or the logistic model of Verhulst (with constraints). 

Economic entities, typically measured in size of revenues, market capitalization, or more rarely by total assets, do not multiply per se but grow themselves by increasing their revenue, profits, and total value through a combination of pricing and market share. Their growth, even in base cases without competition, is never assumed to become exponential and growth rates well below a doubling of 100\% are expected except for the smallest of companies. Also, while biological units live to reproduce into the future, economic units try to exist and grow indefinitely with no need to create progeny.

\item	Steady State: Most economic models in classical and neoclassical economics rely on equilibrium conditions and market clearing to look at long-term steady states in markets. More recent work has introduced multiple equilibria or equilibrium surfaces but this steady state remains as a key long-term assumption in many models. Though Schumpeter and other economists have discussed innovation as driving growth in a system that would otherwise be static, equilibrium is still seen to be the standard, only to be displaced by shocks. In biological systems, there is typically no such steady state. Even in the simplest models of Lotka-Volterra predator prey dynamics, extinction or oscillating limit cycles are seen as valid modes of existence. While biologists are interested in the questions of stability of ecosystems, no equilibrium population or ecology is necessarily expected. This may be in part due to response time lags by biological systems, who rarely can ad hoc innovate to new challenges and must wait for natural selection acting on population variations to adjust populations and species to new environments.

\item	Covariance  and Flow (or feedback) Between Levels: Biological networks have long been known to exhibit an obvious negative covariance between trophic levels as far as population growth is concerned. This is of course due to predation from a higher level which increases the predator population, but reduces the prey population at a lower level. Negative covariance in biological networks is often believed to be a key factor in their stability \cite{biostab}. In economic networks, however, the usual course is for mutual benefit for both partners. If Company A grows, it increases purchases from its suppliers who also grow as well. This positive feedback is distinctly different from biological networks and underlines economic growth. It also may be possible that there are opposite effects for amplified perturbations across levels in both networks where high covariance between predator and prey populations strengthens trophic cascades while high covariance (or cooperation) in business networks damps the bullwhip effect \cite{cascades}. 

In addition, directions of flow that sustain each trophic level are reversed. In biological networks, energy flows through predation to ever higher trophic levels starting from the base autotrophs. Each level is dependent on the availability of food at lower levels. In business networks, demand flows the opposite way down the value chain from end consumers through producers and multiple tiers of suppliers. Each level is dependent on demand at its customers which are typically higher up the value chain.

\item	Clear Species Boundaries: This is related to the problem in the first difference. In the biological world, species boundaries, while not completely agreed upon in many instances, tend to be clearer as the lineage of organisms diverges. In economic systems, where entities can merge and divest at will, you can often find the same entity playing in different market niches while drawing on the same resources such as access to a credit facility or equity investors. This creates two distinctions-first the boundary between competitive entities is often rapidly fluctuating and second it complicates the usual determination of fitness which in biology is amongst members of a population in a species-not across different species. Fitness in economic systems is often defined amongst companies competing in a similar market niche. If fitness is defined across different markets for the same company it is an open question how the overall `fitness' of a company would be defined. For example, if Google were to start baking sourdough bread and selling it in the Bay Area, even at a substantial profit, this venture may have a strong `fitness' in the local bread market, but it is questionable whether this diversion of resources increases a measure of `fitness' for the company overall.

\item	Evolutionary Processes: Amongst biological systems there are generally four main processes which drive evolution - genetic drift, random mutation, natural selection, and gene flow. Natural selection is assumed to act on the random variation generated by mutation and gene flow. Genetic drift removes variation over time. Economic networks are assumed mainly to act on a selection process based on variation that can be generated or eliminated by more teleological processes. Variation is not considered as random but is considered to be unevenly distributed \cite{hayek} and to be generated by a variety of mechanisms including capital and funds availability, skills, social or institutional access, culture, regulatory environment, etc. Selection results not in more offspring but in larger revenues through both market growth and capturing market share. 

\end{enumerate}

Biological and economic networks do share many similarities and comparisons can be useful. However, they must be used when appropriate and tempered by the differences in their actual function in practice.


\begin{thebibliography} {1}

\bibitem{hist1}Penrose, E.T.,  Biological Analogies in the Theory of the Firm, \emph{Am. Econ. Rev.},  \textbf{42} (1952) 804-819.

\bibitem{hist2}Hodgson, G.M. (Ed.), \emph{Economics and Biology}, (Edward Elgar Publishing, Cheltenham, UK, 1995).

\bibitem{hist3}Hodgson, G.M. Darwinism in economics: from analogy to ontology, \emph{J. of Evol. Eco.}, \textbf{12} (2002) 259-281.

\bibitem{hist4}Foster, J., The analytical foundations of evolutionary economics: From biological analogy to economic self-organization, \emph{Struct. Change and Econ. Dyn.}, \textbf{8} (1997) 427-451.

\bibitem{hist5}Foster, J., Competitive selection, self-organisation and Joseph A. Schumpeter, \emph{J. Evo. Econ.}, \textbf{10} (2000) 311-328.

\bibitem{hist6}Foster, J., From simplistic to complex systems in economics, \emph{Camb. J. Econ.}, \textbf{29} (2005) 873-892.

\bibitem{marshall1}Marshall, A. \emph{Principles of Economics}, (MacMillan, London, 1890).

\bibitem{marshall2}Niman, N.B., Biological analogies in Marshall's Work, \emph{J. Hist. Econ. Thought}, \textbf{13} (1991) 19-36.

\bibitem{marshall3}Limoges, C. \& Ménard, C., Organization and the division of labor: biological metaphors at work in Alfred Marshall's Principles of Economics in \emph{Natural images in economic thought: ``markets read in tooth and claw''} edited by Mirowski, P. (Cambridge Univ. Press, Cambridge, UK, 1994) pp. 336-359.

\bibitem{biostab}Tilman, D., Lehman, C.L., Bristow, C.E., Diversity-stability relationships: statistical inevitability or ecological consequence?, \emph{Am. Nat.}, \textbf{151} (1998) 277-282. 

\bibitem{cascades}Smith, R.D., Propagation of Cascades in Complex Networks: From Supply Chains to Food Webs, (available at http://arxiv.org/abs/1103.4983).

\bibitem{hayek}Hayek, F., The use of knowledge in society, \emph{Am. Econ. Rev.}, \textbf{35} (1945) 519-530.

\end{thebibliography}
\end{document}